\documentclass[fleqn,usenatbib]{mnras}

\usepackage{newtxtext,newtxmath}
\usepackage[T1]{fontenc}

\DeclareRobustCommand{\VAN}[3]{#2}
\let\VANthebibliography\thebibliography
\def\thebibliography{\DeclareRobustCommand{\VAN}[3]{##3}\VANthebibliography}

\usepackage{graphicx}	% Including figure files
\usepackage{amsmath}	% Advanced maths commands

\usepackage{cleveref}

\crefformat{figure}{Fig.~#2#1#3}
\crefformat{equation}{equation~(#2#1#3)}
\crefformat{section}{section~#2#1#3}
\crefformat{table}{Tab.~#2#1#3}
\crefformat{appendix}{appendix~#2#1#3}
\crefmultiformat{figure}{Figs.~#2#1#3}{~and~#2#1#3}%
    {,~#2#1#3}{,~#2#1#3}
\crefrangeformat{figure}{Figs.~(#3#1#4--#5#2#6)}
\crefrangeformat{equation}{equation~(#3#1#4--#5#2#6)}

\newcommand{\DJetal}{DJ23}

\title[Posterior Recovery with Emulators]{On the accuracy of posterior recovery with neural network emulators}

\author[H. T. J. Bevins et al.]{
H. T. J. Bevins,$^{1, 2}$\thanks{E-mail: htjb2@cam.ac.uk}
T. Gessey-Jones,$^{1, 2}$\thanks{Now at PhysicsX, Victoria House, 1 Leonard Circus, London, UK, EC2A 4DQ}
W. J. Handley$^{1, 2}$
\\
$^{1}$Kavli Institute for Cosmology, Madingley Road, Cambridge CB3 0HA, UK\\
$^{2}$Astrophysics Group, Cavendish Laboratory, J.J. Thomson Avenue, Cambridge CB3 0HE, UK\\
}

\date{Accepted XXX. Received YYY; in original form ZZZ}

\pubyear{\the\year{}}

\begin{document}
\label{firstpage}
\pagerange{\pageref{firstpage}--\pageref{lastpage}}
\maketitle

% Abstract of the paper
\begin{abstract}
    Neural network emulators are widely used in astrophysics and cosmology to approximate complex simulations inside Bayesian inference loops. Ad hoc rules of thumb are often used to justify the emulator accuracy required for reliable posterior recovery. We provide a theoretically motivated limit on the maximum amount of incorrect information inferred by using an emulator with a given accuracy. Under assumptions of linearity in the model, uncorrelated noise in the data and a Gaussian likelihood function, we demonstrate that the difference between the true underlying posterior and the recovered posterior can be quantified via a Kullback-Leibler divergence. We demonstrate how this limit can be used in the field of 21-cm cosmology by comparing the posteriors recovered when fitting mock data sets generated with the 1D radiative transfer code \textsc{ARES} directly with the simulation code and separately with an emulator. This paper is partly in response to and builds upon recent discussions in the literature which call into question the use of emulators in Bayesian inference pipelines. Upon repeating some aspects of these analyses, we find these concerns quantitatively unjustified, with accurate posterior recovery possible even when the mean RMSE error for the emulator is approximately 20\% of the magnitude of the noise in the data. For the purposes of community reproducibility, we make our analysis code public at this link \url{https://github.com/htjb/validating_posteriors}.
\end{abstract}

% Select between one and six entries from the list of approved keywords.
% Don't make up new ones.

%%%%%%%%%%%%%%%%%%%%%%%%%%%%%%%%%%%%%%%%%%%%%%%%%%

%%%%%%%%%%%%%%%%% BODY OF PAPER %%%%%%%%%%%%%%%%%%

\section{Introduction}

In cosmology and astrophysics, researchers make frequent use of Bayes theorem and Bayesian inference to perform model comparison and parameter estimation. Most astrophysical and cosmological models are complex and computationally expensive, relying on a variety of numerical and semi-numerical simulation techniques. Such simulations are often too expensive to use in inference algorithms, and neural network surrogates or emulators are now commonly used to efficiently approximate these models.

Neural Network emulators have been developed in a variety of fields including SED fitting \citep[e.g.][]{Alsing2020Speculator, Mathews2023Emulators, Mathews2025Emulators}, CMB primary and secondary studies \citep[e.g.][]{Auld2007CosmoNet, Agarwal2014PkANN, Manrique-Yus2020, Albers2019CosmicNet, Arico2021, Mootoovaloo2022, SpurioMancini2022cosmopower, Bonici2024Capse, Bolliet2024Emulators, Gunther2023, Gunther2025}, studies of effective field theory of large scale structure \citep[e.g.][]{Bonici2025effort}, 21-cm cosmology \citep[e.g.][]{Cohen202021cmGEM, Bevins2021globalemu, Bye202221cmVAE, Breitman202421cmEMU, DorigoJones202421cmLSTM} and others \citep[e.g.][]{ElGammal2023Gpry}. They are trained on sets of example simulations from numerical or semi-numerical codes to translate between physical parameters such as the Hubble constant or star formation efficiency and return various summary statistics such as a power spectrum. Often when designing emulators, researchers either try to make the emulators as accurate as possible or alternatively use ad hoc rules of thumb to define minimum requirements for the accuracy. For example, when designing the 21-cm emulator \textsc{globalemu} the authors suggested that the emulator should have an average accuracy $\leq 10\%$ of the expected noise in a 21-cm experiment \citep{Bevins2021globalemu} so that the contribution to the uncertainty from the emulator in any analysis is subdominant to the uncertainty introduced by the instrument.

However, researchers often talk about emulator accuracy in terms of how well the signal can be recovered for a particular set of parameters relative to the underlying simulation that the emulator is trained on. This is typically measured via a (root) mean squared error, which is often an attractive choice because it closely represents the loss function that the network is trained on. What is perhaps more important, however, is how well the posterior can be recovered when using an emulator relative to the true posterior recovered when using the full simulation. This is harder to measure as we typically do not have access to the true posterior, and it is not clear how rules of thumb like those described above translate into posterior bias.

\cite{DorigoJones2023Validation} [hereafter \DJetal] began to discuss this topic within the field of 21-cm cosmology.  Researchers in this field aim to observe the evolution of the 21-cm signal from the spin-flip transition of neutral hydrogen. The signal traces the formation of the first stars and galaxies through to the moment at which the Universe transitioned from being predominantly neutral to ionized. The signal can be observed in the radio band as a sky-averaged signature with a single radiometer, or alternatively as a power spectrum with an interferometer. Although no current confirmed detections of either the power spectrum or the sky-averaged signal exist, there are a number of upper limits on both and emulators have been used extensively to analyse these data sets \citep[e.g.][]{Bevins2022SARAS2, Bevins2022SARAS3, HERA2022Constraints, HERA2023Constraints, EDGESHB2019Constraints}. 

\DJetal{} compares the posteriors recovered when using the 1D radiative transfer code \textsc{ARES} \citep{Mirocha2012ARES, Mirocha2014ARES} and an emulator of \textsc{ARES} built with \textsc{globalemu} \citep{Bevins2021globalemu} to fit a synthetic fiducial signal with varying levels of noise. Running inference tools like MCMC and nested sampling on a likelihood that uses \textsc{ARES} is feasible if computationally intense, as \textsc{ARES} takes of order a few seconds to evaluate. In practice, however, researchers hope to use more detailed semi-numerical simulations in this field to learn about the 21-cm signal, such as 21cmSPACE \citep[e.g.][]{Visbal201221cmSPACE,Fialkov201421cmSPACE, Fialkov201421cmSPACE2,GesseyJones202321cmSPACE, Sikder202421cmSPACE} and 21cmFAST \citep[][]{Mesinger201121cmFAST, Murray202021cmFAST}, and these take hours to run per parameter set. Emulators are therefore crucial in this field, and a host of different tools have been developed \citep{Cohen202021cmGEM, Bevins2021globalemu, Bye202221cmVAE, Breitman202421cmEMU, DorigoJones202421cmLSTM}.

To compare the two posteriors, \DJetal{} defines an emulator bias metric, which quantifies the difference between the means of the 1D posterior distributions in units of the standard deviation of the true 1D distributions. The paper reports that even with a mean emulator error equivalent to 5\% of the expected noise in the data, biased posteriors are recovered. However, several choices are made in preprocessing the training data that are in contrast to the suggestions in the original work \citep{Bevins2021globalemu}. The analysis also includes constraints from UV luminosity functions, which complicates the comparison. The results presented in \DJetal{} have previously been used to justify not using emulators like \textsc{globalemu} \citep[e.g.][]{Saxena2024SBI}.

In this paper, we define an upper bound on the Kullback-Leibler~(KL) divergence between the true and emulated posteriors. We then repeat some of the analysis in \DJetal{} and compare the recovered posteriors using the KL divergence in the context of the defined limit.  We demonstrate that the preprocessing choices made can lead to worse emulation of the sky-averaged 21-cm signal, and show that accurate posteriors can be recovered even with an emulator error of order 20\% of the noise in the data.

In \cref{sec:information-loss} we define the upper limit on the KL divergence between the true and emulated posterior as a function of the noise in the data and emulator accuracy. We then demonstrate this with \textsc{ARES} and \textsc{globalemu} in \cref{sec:example}. We conclude in \cref{sec:conclusions}. Our code is publicly available at \url{https://github.com/htjb/validating_posteriors}.

\section{Measuring information loss}
\label{sec:information-loss}

An emulator $M_\epsilon$ of a model $M$ takes the same parameters $\theta$ as $M$ and gives an output
\begin{equation}
M_\epsilon (\theta) = M(\theta) + \epsilon (\theta),
\end{equation}
where the error $\epsilon(\theta)$ is ideally small for all $\theta$ and 0 in the limit of perfect training.

If an emulator is used in a Bayesian inference pipeline, then the recovered posterior $P_\epsilon (\theta|D, M_\epsilon)$ will differ from the true posterior $P(\theta|D, M)$. For the emulator to be useful then the difference between these two posteriors should be negligible. A natural way to quantify the difference between these two probability distributions is the Kullback-Leibler divergence $\mathcal{D}_\mathrm{KL}$. Generally the $\mathcal{D}_\mathrm{KL}$ between two distributions quantifies the information loss or gain when moving from one to the other. Here the $\mathcal{D}_\mathrm{KL}$ quantifies the incorrect information inferred about the parameter space, in natural bits, when using the emulator to recover the posterior. Ideally, we would like the $\mathcal{D}_\mathrm{KL}$ between these two posterior distributions to be as close to zero as possible and $<1$ so that no significant false information is inferred.

\subsection{The general case}

The error in the emulator $\epsilon$ will induce a corresponding error in the likelihood, $L(\theta) = P(D|\theta, M)$, function
\begin{equation}
    \log L \rightarrow \log L + \delta \log L,
\end{equation}
where we have dropped the $D$, $\theta$ and model $M$ for brevity. The posterior changes as
\begin{equation}
    P(\theta|D, \mathcal{M}) = \frac{L \pi}{\int L \pi d\theta} \rightarrow P_\epsilon(\theta|D, \mathcal{M}_\epsilon) = \frac{L \pi \exp(\delta \log L)}{\int L \pi \exp(\delta \log L) d\theta},
\label{eq:posterior-change}
\end{equation}
where $\pi = P(\theta|M)$ is the prior distribution on the free parameters in the model. We will refer to these posteriors as $P$ and $P_\epsilon$ for brevity.

We can calculate the KL-divergence of this change in posterior as 
\begin{equation}
\mathcal{D}_\mathrm{KL} = \int P \log \left( \frac{P}{P_{\epsilon}} \right) d \theta.
\end{equation}
Substituting in \cref{eq:posterior-change} into the $\log$,
\begin{equation}
\begin{split}
\mathcal{D}_\mathrm{KL} &= \int P \log \left(\exp(-\delta \log L) \frac{\int L \pi \exp(\delta \log L) d \theta}{\int L \pi d\theta} \right) d \theta \\
     &= -\int P \delta \log L d \theta + \int P \log \left( \frac{\int L \pi \exp(\delta \log L) d \theta}{\int L \pi d\theta} \right) d \theta \\
     &= -\langle \delta \log L \rangle_{P} + \int P \log \left( \int P \exp(\delta \log L) d \theta \right) d \theta.
\end{split}
\end{equation}
Recognizing that the term in the log is a constant and thus can be factored out of the integral we find
\begin{equation}
\begin{split}
\mathcal{D}_\mathrm{KL} &= - \langle \delta \log L{} \rangle_{P} + \log \left( \int P \exp(\delta \log L) d \theta \right) \int P d \theta  \\
&= - \langle \delta \log L{} \rangle_{P} + \log \left( \int P \exp(\delta \log L) d \theta \right)  \\
&= \log \left( \frac{\int P \exp(\delta \log L) d \theta}{\exp\left(\int P \delta \log L{} d \theta\right)} \right). 
\label{eq:kl-divergence-general}
\end{split}
\end{equation}
We can also express this result in terms of posterior moments of $\delta \log L$. Defining,
\begin{equation}
\lambda_{n} \equiv \frac{1}{n!} \int P \left(\delta \log L\right)^n d \theta
\end{equation}
and using a Taylor expansion of the exponential in \cref{eq:kl-divergence-general} we can write the KL divergence as
\begin{equation}
\mathcal{D}_\mathrm{KL} = \log\left(\sum_{i = 0}^{\infty} \lambda_{i}\right) - \lambda_{1} = \log\left(1 + \sum_{i = 1}^{\infty} \lambda_{i}\right) - \lambda_{1}.
\end{equation}

In the limit that $\delta \log L$ is small, we can consider the low order expansion of $\mathcal{D}_\mathrm{KL}$. Noting that $\lambda_{n} = O(\delta \log L{}^n)$ and assuming $\left|  \sum_{i = 1}^{\infty} \lambda_{i}\right| <1$ then
\begin{equation}
\begin{split}
\mathcal{D}_\mathrm{KL} &\approx -\lambda_{1} + \left( \sum_{i = 1}^{\infty} \lambda_{i} \right) - \frac{1}{2}  \left( \sum_{i = 1}^{\infty} \lambda_{i} \right) ^2 +  \frac{1}{3}\left( \sum_{i = 1}^{\infty} \lambda_{i} \right)^3 - \ldots \\
&= \lambda_{2} - \frac{1}{2} \lambda_{1}^2 + O\left(\delta \log L{}^3\right),
\end{split}
\end{equation}
Hence, to lowest order in $\delta \log L$ we find that the KL divergence is half of the standard deviation of $\delta \log L$ across the model posterior
\begin{equation}
\mathcal{D}_\mathrm{KL} \approx \frac{1}{2} \left(\langle \delta \log L{}^2 \rangle_{P} - \langle \delta \log L \rangle_{P}^2 \right).
\end{equation} 
While mathematically satisfying we generally do not have access to samples on the true posterior $P$ however we can make progress if we consider the case of a linear model and Gaussian likelihood function.

\subsection{The linear model case}

We denote a linear model as
\begin{equation}
\mathcal{M}(\theta) = M \theta + m
\label{eq:linear-model}
\end{equation} 
and the linear emulator error on this model as
\begin{equation}
\epsilon(\theta) = E \theta + \epsilon
\end{equation}
such that
\begin{equation}
\mathcal{M}_\epsilon (\theta) = (M + E) \theta + (m + \epsilon),
\label{eq:emulator-linear-model}
\end{equation}
where $M$ and $E$ are matrices of dimensions $N_d \times N_\theta$ and $m$ and $\epsilon$ are vectors of length $N_d$. $N_d$ is the number of measured data points and $N_\theta$ is the number of model parameters.

We assume that our prior is uniform in $\theta$ and broad enough that we can ignore the fact a uniform prior is only non-zero in a finite region. We then define our likelihood to be Gaussian
\begin{equation}
L \propto \exp\left(-\frac{1}{2}(D - \mathcal{M})^T \Sigma^{-1} (D - \mathcal{M}) \right),
\end{equation}
where $\Sigma$ is the data covariance matrix.
Substituting in the linear model in \cref{eq:linear-model} and accounting for our assumed prior we get the following Gaussian posterior distribution
\begin{equation}
P \propto \exp\left(-\frac{1}{2}(D -m - M \theta)^T \Sigma^{-1} (D - m - M \theta) \right).
\label{eq:posterior}
\end{equation}
By expanding the above and finding the quadratic term we can see that it has a parameter covariance matrix $C$ given by
\begin{equation}
C^{-1} \equiv M^T \Sigma^{-1} M.
\end{equation}
The mean of the Gaussian $\mu$ can be found by taking the derivative of the expression in the exponential to find its turning point (the posterior maximum)
\begin{equation}
0 = M^T \Sigma^{-1}(D -m - M \mu), 
\end{equation}
where we used the symmetry of $\Sigma$ to combine the two terms. Rearranging we find
\begin{equation}
\begin{split}
M^T \Sigma^{-1} M \mu &= M^T \Sigma^{-1}(D -m) \\
C^{-1} \mu &= M^T \Sigma^{-1}(D -m) \\
\mu &= C  M^T\Sigma^{-1}(D -m).
\end{split}
\end{equation}

Hence, for our linear model we have a Gaussian parameter posterior with covariance
\begin{equation}
C = \left(M^T  \Sigma^{-1} M \right)^{-1},
\label{eq:cov-model}
\end{equation}
and mean
\begin{equation}
\mu = C  M^T\Sigma^{-1}(D -m),
\label{eq:mean-model}
\end{equation}
with the equivalents for the emulator analogously found using \cref{eq:emulator-linear-model} as
\begin{equation}
C_{\epsilon} = \left((M + E)^T  \Sigma^{-1} (M + E) \right)^{-1},
\label{eq:cov-emu}
\end{equation}
and mean
\begin{equation}
\mu_{\epsilon} = C_{\epsilon}  (M + E)^T\Sigma^{-1}(D -m - \epsilon).
\label{eq:mean-emu}
\end{equation}
These are all standard results \citep{MatrixCookbook}.

To work out $\mathcal{D}_\mathrm{KL}$ between these posteriors, we thus need to know the $\mathcal{D}_\mathrm{KL}$ between two multivariate Gaussians.
This is a well known result and is given by
\begin{equation}
\begin{aligned}
\mathcal{D}_\mathrm{KL} = \frac{1}{2} \bigg[\log\left(\frac{|C_{\epsilon}|}{|C|}\right) & - N_\theta + {\rm tr}\left(C_{\epsilon}^{-1} C\right) + \\ &(\mu_{\epsilon} - \mu)^T C_{\epsilon}^{-1}(\mu_{\epsilon} - \mu)  \bigg].
\end{aligned}
\label{eq:multivariate_gauss_kl_general}
\end{equation}
The above set of five \crefrange{eq:cov-model}{eq:multivariate_gauss_kl_general} are the general solution for the KL divergence between $P$ and $P_\epsilon$ for a linear model and Gaussian likelihood.

\subsection{White noise and E=0}

In many cases, the emulator error may evolve slowly over the parameter space $E \ll M$ compared to the model.
We can then approximate $E + M \approx M$ and so
\begin{equation}
    C_{\epsilon} \approx C,
\end{equation}
and
\begin{equation}
    \mu_{\epsilon} \approx \mu - CM^T\Sigma^{-1} \epsilon,
\end{equation}
Substituting these results into the $\mathcal{D}_\mathrm{KL}$ equation (and using symmetry of $\Sigma$ and $C$) it simplifies to 
\begin{equation}
\begin{split}
\mathcal{D}_\mathrm{KL} &= \frac{1}{2} \left[\log\left(1\right) - N_{\theta} + {\rm tr}\left(\mathbf{1}_{N_\theta}\right) +  \epsilon^T\Sigma^{-1} MC C^{-1}CM^T\Sigma^{-1} \epsilon  \right],\\
&= \frac{1}{2} \left[0 - N_{\theta} +N_{\theta}  +  \epsilon^T\Sigma^{-1} MC M^T\Sigma^{-1} \epsilon  \right],\\
&= \frac{1}{2}   \epsilon^T\Sigma^{-1} MC M^T\Sigma^{-1} \epsilon, \\
&= \frac{1}{2}   \epsilon^T\Sigma^{-1} M \left(M^T  \Sigma^{-1} M \right)^{-1} M^T\Sigma^{-1} \epsilon,
\end{split}
\end{equation}
which is a more compact expression that has no $m$ dependence. We see clearly here that $\mathcal{D}_\mathrm{KL}$ is as a quadratic measure on $\epsilon$.

For white noise in the data $\Sigma = \frac{1}{\sigma^2} \mathbf{1}_{N_{\rm d}}$. The above then simplifies further to
\begin{equation}
    \mathcal{D}_\mathrm{KL}  = \frac{1}{2}   \frac{1}{\sigma^2}\epsilon^T M \left(M^T  M \right)^{-1} M^T\epsilon =  \frac{1}{2}   \frac{1}{\sigma^2}\epsilon^T M M^+ \epsilon.
\end{equation}
where $M^{+}$ is the Moore-Penrose inverse of $M$. we can see that the accuracy of the recovered posterior $P_\epsilon$ is dependent on the magnitude of the noise in the data and how sensitive the model $\mathcal{M}$ is to the parameters $\theta$. $MM^+$ is positive definite, and so we are assured that this inner product is positive unless $\epsilon = 0$. Emulator error therefore always introduces some inaccuracy in the posterior, as we would expect.  

As $MM^+$ is real positive definite, its eigenvalues $\lambda_{\rm n}$ are all $\geq 0$ and there exists a real rotation matrix $U$ that rotates us into the orthogonal eigenbasis of $MM^+$, so that $U^{-1}MM^+U = \rm{diag}(\lambda_{\rm n})$.
Note $U$ is a rotation matrix so $U^{-1} = U^T$, with $U^T$ also a rotation matrix, the inverse rotation.
Inserting this transform into the above we find 
\begin{equation}
    \mathcal{D}_\mathrm{KL} =  \frac{1}{2}   \frac{1}{\sigma^2}\varepsilon^T U \rm{diag}(\lambda_{\rm n}) U^{-1} \varepsilon.
\end{equation}
Calling the rotated vector $f = U^{-1} \varepsilon$, we are left with an inner product over a diagonal matrix of positive values
\begin{equation}
     \mathcal{D}_\mathrm{KL} =  \frac{1}{2}   \frac{1}{\sigma^2} \sum_j f_j^2 \lambda_{\rm j}.
\end{equation}
Hence as this is a sum of the product of non-negative values
\begin{equation}
    \mathcal{D}_\mathrm{KL} \leq  \frac{1}{2}  \frac{1}{\sigma^2} \max(\lambda_{\rm n}) \sum_j f_j^2 = \frac{1}{2}  \frac{1}{\sigma^2} \max(\lambda_{\rm n}) ||f||^2.
\end{equation}
But, since $U^{-1}$ is a rotation matrix it leaves the 2-norm of the vector it acts on unchanged, hence $||f|| = ||\epsilon||$ and so
\begin{equation}
     \mathcal{D}_\mathrm{KL} \leq   \frac{1}{2}  \frac{1}{\sigma^2} \max(\lambda_{\rm n}) ||\epsilon||^2.
\end{equation}

The matrix $M M^+$ is a projection matrix since it is idempotent $M M^+ M M^+ = M M^+$. As a result, all of the eigenvalues of $M M^+$ are, in fact, either 1 or 0. Thus $\max(\lambda_{\rm n})  = 1$ and substituting $|\epsilon|^2 = N_{\rm d} \mathrm{RMSE}^2$ we find
\begin{equation}
    \mathcal{D}_\mathrm{KL}  \leq   \frac{N_{\rm d}}{2}  \left(\frac{\rm RMSE}{\sigma}\right)^2,
    \label{eq:limit-dkl}
\end{equation}
where RMSE is the root mean squared error across a test data set for the emulator. Therefore, under the assumptions of Gaussian noise and a linear model for the data, we can say that for less than 1 bit of difference\footnote{Intuitively, 1 bit of information is the amount gained when flipping a fair coin, since there are two equally likely outcomes of this process. Bits can be thought of as the number of yes/no questions that need to be asked to learn the true answer. Note KL divergence is often calculated in nat bits (base $e$) however this is trivially related to the base 2 KL divergence by $\mathcal{D}_\mathrm{KL}^\mathrm{nat} = \mathcal{D}_\mathrm{KL}^\mathrm{bits} \log(2)$.} between $P$ and $P_\epsilon$ then
\begin{equation}
    1 \leq \frac{N_{\rm d}}{2}  \left(\frac{\rm RMSE}{\sigma}\right)^2,
\end{equation}
which when inverted gives
\begin{equation}
    \frac{\rm RMSE}{\sigma}  \leq \sqrt{\frac{2}{N_{\rm d}}}.
    \label{eq:limit}
\end{equation}
When using an emulator, we see that to maintain the same bound on inferred inaccurate information as the number of data points increase, the emulator accuracy needs to improve. This is intuitively quite satisfying since, for independent and identically distributed random variables, having more data points should give more tightly constrained posteriors and so model accuracy needs to be higher.  For $N_d \sim 100$, typical of the expected number of data points in a 21-cm global signal observation, then
\begin{equation}
     \frac{\rm RMSE}{\sigma}  \leq 0.14,
\end{equation}
meaning that for less than one bit of difference between $P$ and $P_\epsilon$ then the average RMSE across the independent variable for any emulator should be less than 14\% of the experimental noise. This interestingly agrees well with the intuition used in \cite{Bevins2021globalemu} to justify the required level of accuracy of a 21-cm emulator.

\subsection{Validity of the assumptions made}

In the above derivation there are three key assumptions that are being made; the model and emulated model can be approximated by a linear model, the likelihood is Gaussian and the noise in the data is uncorrelated. 

The sum of many identically distributed random variables tends towards a normal distribution, regardless of the original distribution of each variable. This means that in the limit of large amounts of data then the likelihood and hence posterior, in the weakly informative prior case, tend towards Gaussian distributions. Taking the first order Taylor expansion of the model around the peak of a Gaussian like posterior distribution often results in near linearity in the model because the curvature of the distribution is small. The assumption breaks down when there is limited data or when we move far away from the posterior peak, but since the $\mathcal{D}_\mathrm{KL}$ is an average over the posterior the behaviour in the tails of the distribution will have less impact on its value. It is also only true locally in the case of multi-modal posteriors, but these are arguably rare in cosmology.

As discussed in the introduction, this work was largely motivated by the use of neural network emulators in the field of 21-cm cosmology. In this field it is common to use a Gaussian likelihood function \citep{Anstey2021REACH, Scheutwinkel2023Likelihoods} with radiometric noise given by
\begin{equation}
    \sigma(\nu) = \frac{T_{A}(\nu)}{\sqrt{\Delta \nu \tau}},
    \label{eq:radiometric-noise}
\end{equation}
where $\nu$ is frequency of the observations, $T_A$ is the antenna temperature, $\Delta \nu$ is the channel width of the data and $\tau$ is the integration time for the observations. In a narrow bandwidth $\sigma(\nu)$ is approximately constant and most studies currently assume a constant level of noise across the observed frequency range. However, $\sigma$ can be replaced in \cref{eq:limit} with $\mathrm{min}(\sigma(\nu))$.

\cite{Scheutwinkel2023Likelihoods} explored the use of other likelihood functions in the field of 21-cm cosmology. Since we are assuming the prior is uniform, then the posterior in \cref{eq:posterior} is approximately equivalent to the likelihood function. If an analytic expression exists for the $\mathcal{D}_\mathrm{KL}$ between the two approximate posterior distributions, equivalent to \cref{eq:kl-divergence-general}, then similar arguments to those proposed here can be followed to arrive at an upper bound on the KL divergence between $P$ and $P_\epsilon$.

We note that in higher dimensional spaces, the assumption that the model is linear around the peak of the posterior can become less valid because the posteriors are more complex (increased curvature and/or multimodal structure). Assuming linearity does hold around the peak of the posterior, most of the posterior mass may lie far from the peak, meaning that the KL divergence is dominated by regions in which linearity fails. Although, the limit is not directly dependent on the number of parameters it is dependent on the complexity of the posterior, which is a function of the dimensionality of the model. A full analysis of this behaviour is beyond the scope of this paper and left for future work.

\section{An example in 21-cm Cosmology}
\label{sec:example}

\subsection{21-cm cosmology}

We now illustrate the utility of the above using an example from the field of 21-cm cosmology. The 21-cm signal from neutral hydrogen during the cosmic dawn and epoch of reionization is a powerful probe of the properties of the first stars and the intergalactic medium (IGM) at high redshifts. The signal originates from the spin-flip transition in the neutral hydrogen, and the relative number of atoms with aligned and anti-aligned electron and proton spins is characterised by a statistical temperature. The rate of the spin-flip transition is driven by interactions between the neutral hydrogen, the cosmic microwave background, light from the first stars and the kinetic temperature of the gas that makes up the IGM \citep{Furlanetto2006Review, Barkana2016Review, Mesinger2019Review}.

Numerical, Semi-numerical, 1D radiative transfer codes and analytic models are all used to model the evolution of the 21-cm signal and parameterize its dependence on the properties of the first stars and galaxies. Numerical hydro simulations like C2-Ray while detailed are extremely computationally expensive \citep{Mellema2006C2RAY}. 1D radiative transfer codes like ARES \cite{Mirocha2012ARES, Mirocha2014ARES}, Zeus21 \citep{Munoz2023zeus21} and ECHO21 \citep{Mittal2025echo21} offer a computationally cheaper alternative, but are not as detailed as numerical models. Semi-numerical codes like 21cmSPACE and 21cmFAST can generally be evaluated in a few hours and offer a relatively cheap compromise somewhere between full hydro simulations and 1D codes, allowing for a wider range of physical processes to be modelled.

The signal is measured relative to the radio background and is redshifted into the radio band. Observers are attempting to detect the sky-averaged evolution of this signal over time with single antennas \citep{EDGES, MIST, PRIZM, LEDA, REACH, PRATUSH, RHINO, SARAS3, LCRT, RHINO,LUSEE} and the spatial fluctuations via the power spectrum \citep{HERA,MWA, LOFAR, NenuFAR,ALO, SKA,DSL, FarView}.

Several interferometers have placed upper limits on the magnitude of the power spectrum and in 2018 a tentative detection of the sky-averaged signal was made by the EDGES collaboration \citep{EDGES} although this was recently disputed by observations from the SARAS3 instrument \citep{Singh2022SARAS3}. Previous concerns have also been raised about the cosmological origins of the EDGES signal \citep{Hills2018ConcernsAboutEDGES, Sims2020ConcernsAboutEDGES, Singh2019ConcernsAboutEDGES, Bradley2019ConcernsAboutEDGES, Bevins2021ConcernsAboutEDGES}.

A number of works \citep[e.g.][]{EDGESHB2019Constraints, HERA2022Constraints, SARAS22018Constraints, HERA2023Constraints, Bevins2022SARAS3, Bevins2022SARAS2, Pochinda2024constraints, GesseyJones2024constraints, Bevins2024Constraints} have used upper limits on the magnitude of the sky-averaged or global 21-cm signal and the power spectrum to constrain the properties of the first stars and galaxies. Since 1D radiative transfer codes can be evaluated on the order of seconds, they can be used in Bayesian inference pipelines. However, to make use of semi-numerical and numerical codes, researchers rely on neural network emulators because it is computationally infeasible to run inference directly on the semi-numerical simulations which take of order hours to evaluate per parameter set. Several emulator frameworks of the 21-cm signal exist, such as \textsc{21cmVAE} \citep{Bye202221cmVAE}, \textsc{21cmGEM} \citep{Cohen202021cmGEM} and \textsc{21cmEMU} \citep{Breitman202421cmEMU}. In this work, we compare the posteriors recovered when using the 1D radiative transfer code ARES and an emulator of ARES built with \textsc{globalemu} to fit mock data of varying noise levels.

\subsection{Previous work}

We now summarise the analysis presented in \DJetal{}. First, a data set of 21-cm global signal simulations were generated using the ARES code. \textsc{globalemu} emulators were then trained on these simulations. Using these emulators as well as the ARES code directly, inference was performed on synthetic global 21-cm signal data with various noise levels, both in isolation and jointly with UV luminosity function (UVLF). The resulting 1D posteriors were then compared using two different metrics
\begin{equation}
    \mathrm{emulator~bias} = \frac{|\mu_{\textsc{globalemu}} - \mu_{\textsc{ARES}}|}{\sigma_{\textsc{ARES}}}
    \label{eq:emulator-bias}
\end{equation}
and
\begin{equation}
    \mathrm{true~bias} = \frac{|\mu_{\textsc{ARES}} - \theta_0|}{\sigma_{\textsc{ARES}}}
\end{equation}
where $\mu_{\textsc{globalemu}}$ and $\mu_{\textsc{ARES}}$ are the means of the posteriors for each parameter, $\sigma_{\textsc{ARES}}$ is the standard deviation for the 1D \textsc{ARES} posteriors and $\theta_0$ are the true values of the parameters that were used to generate the mock data. 

While interesting, the `true bias' is not useful for validating the recovery of the posterior with $\textsc{globalemu}$.  The `true bias' can be very large, indicating a poor fit to the data, but if the `emulator bias' is zero then we would conclude that using the emulator has not introduced any additional uncertainty into the analysis. What matters most for the purposes of this work is whether the two posteriors are the same, not whether the true parameters are recovered. The `emulator bias' does go some way towards answering this question, although it has some limitations. The metric quantifies how far apart the means of the 1D posteriors are from each other in units of standard deviation of the ARES posterior. However, as it only compares the 1D posteriors, it does not take into consideration higher dimensional differences. Additionally, it only tells you whether the distributions are centred in the correct place, not if the emulator posterior is over or under confident\footnote{For some examples of non-trivial 2D distributions with equivalent means and standard deviations on each parameter see the Datasaurus data set at \url{https://en.wikipedia.org/wiki/Datasaurus_dozen}.}. \DJetal{} includes a discussion of using a Kolmogorov-Smirnov test in appendix A, which like the `emulator bias' only compares the 1D marginalised posteriors (an N-dimensional version of the Kolmogorov-Smirnov test was formulated in \cite{Harrison2015NDKSTest}).

\textsc{globalemu} includes several physically motivated preprocessing steps to improve the accuracy of trained emulators. These were switched off in \DJetal{}. The paper claims that this makes the emulators more accurate, in contrast to the conclusions in the original \textsc{globalemu} paper. In \cref{sec:emulator}, we show that more accurate emulators can be trained with the preprocessing steps on using the training data from \DJetal{}.

The comparison between the posteriors recovered with the two approaches in \DJetal{} is more difficult due to the inclusion of UV luminosity constraints. The paper generates mock luminosity functions using \textsc{ARES} with the same set of parameters used for the sky-averaged 21-cm signal data and calibrated this to observations from \cite{Bouwens2015UVLF}. The paper then performs joint analysis of both the UV luminosity function and the 21-cm signal by fitting both with \textsc{ARES} and separately the 21-cm signal with \textsc{globalemu} and the UV luminosity function with \textsc{ARES}. To allow for a more direct comparison between the two sets of posteriors, we do not include UV luminosity constraints in this work.

\subsection{\textsc{ARES} setup}

We use the same training and test data sets from \DJetal{}. We refer the reader to that work and references therein for more detail on the parameterisation of \textsc{ARES}, but briefly summarise the modelling below. There are eight free parameters in the model governing different physical processes.

The X-ray efficiencies of galaxies is governed by a normalising factor, $c_x$, on the X-ray luminosity-star formation rate relationship and $\log N_\mathrm{HI}$ is the neutral hydrogen column density in galaxies.

The star formation efficiency~(SFE) is parameterised by a double power law function
\begin{equation}
    f_*(M_\mathrm{h}) = \frac{f_*}{\bigg(\frac{M_\mathrm{h}}{M_\mathrm{c}}\bigg)^{\gamma_\mathrm{lo}} + \bigg(\frac{M_\mathrm{h}}{M_\mathrm{c}}\bigg)^{\gamma_\mathrm{hi}} },
\end{equation}
where $M_\mathrm{h}$ is the halo mass, $f_*$ is two times the star formation efficiency at a halo mass of $M_\mathrm{c}$ and $\gamma_\mathrm{lo}$ and $\gamma_\mathrm{hi}$ are the slopes of the low and high mass ends of the SFE. The double power law is motivated by observations of the galaxy and halo mass functions and the expected suppression of star formation in large halos from various feedback mechanisms.

A high absolute value of $\gamma_\mathrm{lo}$ leads to a suppression of star formation in low mass halos, which delays the onset of the Cosmic Dawn. A high value of $\gamma_\mathrm{hi}$ means that there is strong feedback in large galaxies that suppresses star formation. Since larger galaxies will not form until much later in cosmic history the global 21-cm signal is largely insensitive to $\gamma_\mathrm{hi}$ but because it gives us information about the timing of the Cosmic Dawn it is very sensitive to $\gamma_\mathrm{lo}$. As discussed in \DJetal{} we need other probes to constrain the high mass end of the star formation efficiency such as UV Luminosity Functions. The 21-cm signal is also quite sensitive to the value of $T_\mathrm{min}$, the minimum virial temperature for star-forming halos, as a large temperate can suppress star formation. 

Finally, the escape fraction of UV photons which ionize the neutral hydrogen is parametrised as $f_\mathrm{esc}$. The value of $f_\mathrm{esc}$ controls how quickly the Universe reionizes, and thus how quickly the 21-cm signal disappears. A high $f_\mathrm{esc}$ leads to a rapid reionization and a 21-cm signal that quickly vanishes.

We use the same prior ranges as in \DJetal{} on our parameters and the same fiducial values for our ground truth model (see \cref{tab:fiducial-parameter-set}) to which we add Gaussian distributed noise with a standard deviation of 5, 25, 50 and 250 mK in our experiments.

\begin{table}
    \centering
    \begin{tabular}{|c|c|}
        Parameter & Value \\
        \hline
        $f_\mathrm{esc}$ & 0.2 \\
        $c_x$ & $2 \times 10^{39}$ erg s$^{-1}$(M$_\odot$ yr$^{-1}$)$^{-1}$\\
        $T_\mathrm{min}$ & $10^{4}$ K \\
        $\log N_\mathrm{HI}$ & 21 \\
        $f_*$ & 0.05 \\
        $M_c$ & $2\times 10^{11}$ M$_\odot$ \\
        $\gamma_\mathrm{lo}$ & 0.49 \\
        $\gamma_\mathrm{hi}$ & -0.61 \\
    \end{tabular}
    \caption{The fiducial parameter set for the mock data analysed in this paper and \DJetal{}.\label{tab:fiducial-parameter-set}}
\end{table}

\subsection{\textsc{globalemu} emulator}
\label{sec:emulator}

\textsc{globalemu} is a flexible tool for building 21-cm emulators and has a number of optional preprocessing steps built in. These steps can be turned off, but their application is recommended, as they are designed to make the problem easier for the network to learn and to emphasize the differences in the 21-cm signal corresponding to different astrophysical models. These preprocessing steps are detailed in \cite{Bevins2021globalemu}. We briefly recap the preprocessing below and demonstrate that when emulating the \textsc{ARES} data set used in \DJetal{} this preprocessing can lead to an improved performance, contrary to conclusions in that paper.

The first preprocessing step is the subtraction of the Astrophysics Free Baseline~(AFB). The AFB is an approximation to the 21-cm signal during the dark ages, when the signal is largely independent of astrophysics and dominated by cosmology. The AFB is common to all the signals in the \textsc{ARES} training data, and it decreases with increasing redshift. By subtracting the AFB from the training data, we are preventing the network from having to learn a non-trivial but known relationship between the brightness temperature and redshift at high redshift, thus making the problem easier to emulate. 

After subtraction of the AFB, the signals in the training data are resampled along the redshift axis. The \textsc{ARES} training data is sampled uniformly in redshift, but there is no significance to this, and the resampling step increases the redshift sampling where the training data varies most to emphasize this variation to the network.

In \cref{tab:ARES-training-accuracy} we report the accuracy of the emulation when switching on and off the preprocessing steps discussed above. In the original \textsc{globalemu} paper, the accuracy of the emulator was assessed using an unseen test data set and over fitting was checked for after training by comparing the distribution of the error in predicted signals for the test and training data. More recent iterations of the \textsc{globalemu} code implemented a form of early stopping algorithm using the test data set provided to the emulator. In several works, \citep[e.g.][]{Bevins2022SARAS2, Bevins2022SARAS3, Pochinda2024constraints, GesseyJones2024constraints} including \DJetal{} the same test data set used for early stopping was used to assess the accuracy of the emulator. In practice, this is not a fair representation of the performance of the emulator because the emulator has been fine-tuned to perform well on that specific data set. In this work, we generate a new set of simulations with \textsc{ARES} to measure the accuracy of the network.

We train six different versions of the emulator for each combination of preprocessing steps. The accuracy for each was then evaluated over a test data set comprising 2000 models over the band $z=6 - 55$ and the mean Root Mean Squared Error~(RMSE), the 95$^\mathrm{th}$ percentile error and the worst emulation error were calculated. For each combination of preprocessing steps, we then took an average of the recorded mean RMSE values along with a corresponding standard deviation to assess the relative performance of the different emulator set-ups across the stochastic initialisation of the network weights. We did the same for the 95th percentile and worst RMSE values and these are reported in \cref{tab:ARES-training-accuracy}. The emulators are trained on 24,000 simulations covering the prior range outlined in \DJetal{} and early stopping is performed on a different validation set of 2000 signals. We find that the best mean and 95$^\mathrm{th}$ percentile values are recovered when both the AFB subtraction and the resampling are switched on.

In addition to the above preprocessing steps, the training data is scaled by the standard deviation of the signals so that it is of order unity and the parameters $c_X$, $T_\mathrm{min}$, $f_*$ and $M_c$ are all logged.

\begin{table*}
    \centering
    \begin{tabular}{|c|c|c|c|c|c|c|}
    \hline
    Metric & With AFB,& Resampling only & AFB only & No AFB, & Emulator Used & \DJetal{} \\
         & with resampling &  & & no resampling & (With AFB, & No AFB, \\
         & & & & & with resampling) & no resampling \\
         \hline
         Mean & $1.23 \pm 0.16$ & $1.61 \pm 0.16$ & $1.41 \pm 0.12$ & $1.36 \pm 0.08$ & 0.99 & 1.25\\
         95$^\mathrm{th}$ Percentile & $3.96 \pm 0.72$ & $3.98 \pm 0.57$ & $4.59 \pm 0.53$ &  $4.48 \pm 0.33$ &  3.14 & --- \\
         Worst & $31.04 \pm 7.09$ & $31.58 \pm 9.48$ & $33.91 \pm 6.13$ & $33.89 \pm 4.70 $ &  25.97 & 18.5 \\
    \end{tabular}
    \caption{A comparison of emulator accuracy over the band $z=6-55$ as a function of preprocessing steps. We train six different emulators and assess the accuracy of each on 2000 previously unseen signal models for each combination of preprocessing steps. We report the average values over the six training runs and use the standard deviation to show the stochastic variation in the accuracy. While there is some overlap in the performance for the different set-ups, we find that it is possible to train more accurate emulators with the two preprocessing steps switched on, in agreement with the findings in \protect\cite{Bevins2021globalemu}. The most accurate of the emulators that we trained is the one that we use throughout the rest of the paper, and has a mean error of 0.99 mK over the test data set. The values for the \DJetal{} emulator were taken from their paper. 95 \% of the training data have an RMSE lower or equal to the $95^\mathrm{th}$ percentile.}
    \label{tab:ARES-training-accuracy}
\end{table*}

We use the same network architecture as \DJetal{}, namely three hidden layers of 32 nodes each, however as previously noted the paper finds a better performance with the AFB subtraction and resampling switched off. We report in \cref{tab:ARES-training-accuracy} the accuracy of \DJetal{} emulator for ease of comparison.

In this paper, we use version 1.8.0 of \textsc{globalemu} which includes an updated early stopping algorithm compared to that used in \DJetal{}. The previous early stopping algorithm used in \DJetal{} terminates when the training loss does not improve by $10^{-5}$ within the last twenty epochs. However, in version 1.8.0 the early stopping algorithm terminates when the validation loss has not decreased for 10 epochs (2\% of the requested maximum number of epochs) and rolls back to the optimum model.

\begin{figure*}
    \centering
    \includegraphics[]{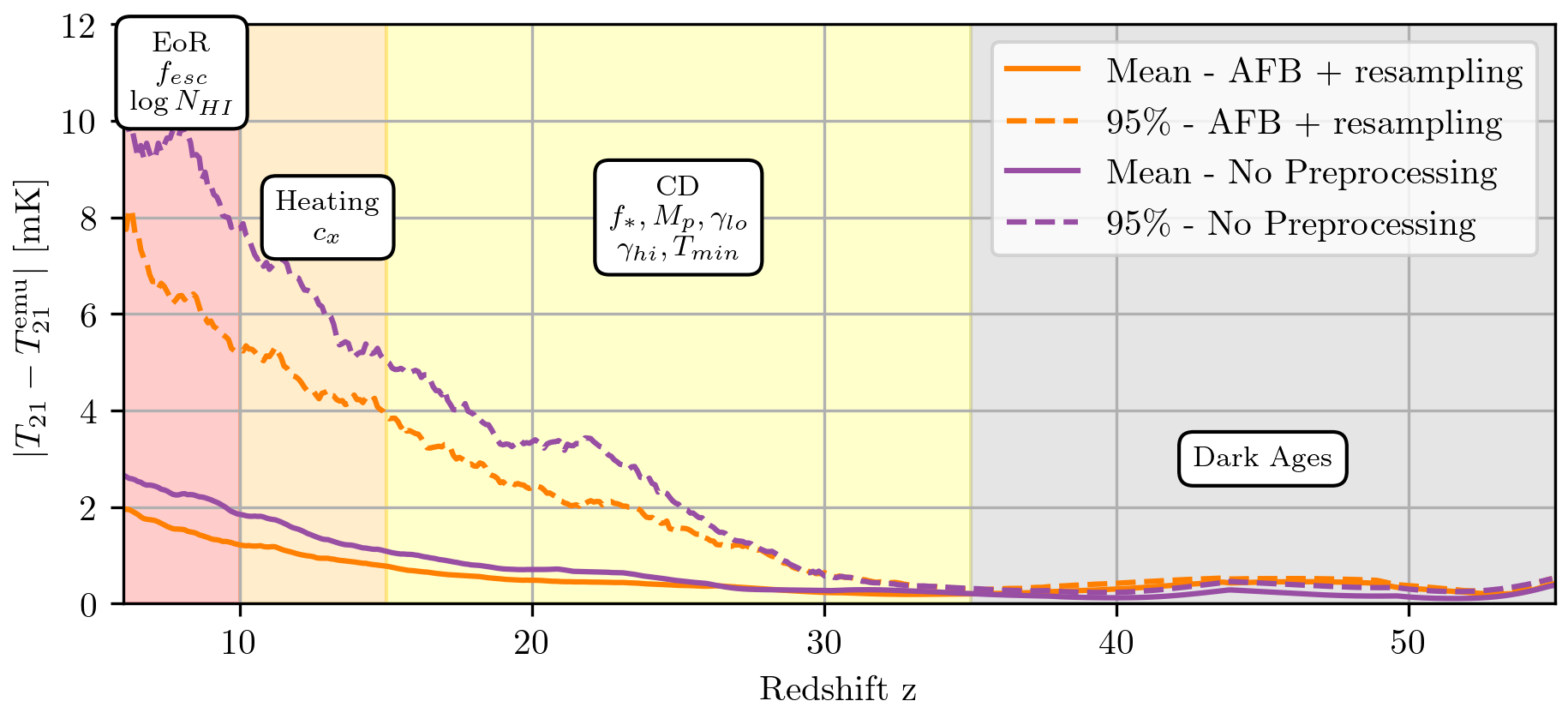}
    \caption{The figure shows the average and 95\% absolute difference between the \textsc{ARES} signals and emulated signals in the test data set as a function of redshift. We show the errors for two emulators, one with the preprocessing steps outlined in the original \textsc{globalemu} paper and another without these preprocessing steps, as was done in \DJetal{}. From the figure we can see that the emulator error is larger at lower redshifts where the variation in the signal across the test data is strongest. We can also see that when we do not include the preprocessing steps, the performance of the emulator is worse, in disagreement with \DJetal{}. The rough redshift ranges corresponding to the Dark Ages (grey), Cosmic Dawn (yellow), Epoch of Heating (orange) and Epoch of Reionization (red) are highlighted along with the set of parameters which have the most impact during each epoch. We note that in reality there is a lot of overlap between these different epochs and the processes that govern the signal. We might expect the recovery of constraints on $f_\mathrm{esc}$ and $\log N_\mathrm{HI}$ to be worse than the constraints on the star formation parameters when using the emulators because the error is larger over the EoR window compared to the CD.\label{fig:emulator-accuracy}}
\end{figure*}

In the analysis that follows, we use the best performing emulator trained with AFB and resampling switched on. The accuracy of the emulator as a function of redshift is shown in \cref{fig:emulator-accuracy} evaluated on the unseen test data.

\subsection{Likelihood functions}

We assume a Gaussian likelihood of the form
\begin{equation}
    \log L = -\frac{N_d}{2} \log (2\pi \sigma^2) - \sum_i^{N_d} \frac{1}{2} \frac{(D_i - \mathcal{M}_i)^2}{\sigma^2},
\end{equation}
where $N_d$ is the number of data points, $D$ is the mock data containing the fiducial \textsc{ARES} signal and Gaussian random noise with standard deviation $\sigma$, $\mathcal{M}$ is the model of the 21-cm signal from \textsc{ARES} or the emulated version from $\textsc{globalemu}$ \footnote{The log-evidences found for the fits in this paper differ to those found in \DJetal{} by around 2000 log units. We do not report them here but they can be found with the chains and the code on github. We believe the difference is because we included the normalisation term on our log-likelihood, $-\frac{N_d}{2} \log (2\pi\sigma^2)$, and \DJetal{} did not. For $N_d = 490$ and $\sigma=25$ mK then this term is equal to approximately $-2027$. }. We are assuming that the errors are not correlated across frequency, which is a common assumption in the field and was made in \DJetal{}.

Usually, the magnitude of the noise in ones' data is not known. Theoretical assumptions can be made about the form and magnitude of the noise using \cref{eq:radiometric-noise}, however, we often fit $\sigma$ as a free parameter in the analysis.
Recent works \citep{GesseyJones2024constraints, Pochinda2024constraints} have included an additional emulator error term in their likelihood functions, such that $\sigma^2 = \sigma_\mathrm{instrument}^2 + \sigma_\mathrm{emulator}^2$. The emulator error is often estimated from the test data sets and fixed in the analysis, whereas the instrument error is fitted for. For a sky-averaged 21-cm experiment, the instrument noise is expected to be around 25 mK \citep[e.g.][]{EDGES, REACH} and emulator errors are typically around 1 mK and therefore $\sigma$ is generally dominated by the contribution from the instrument.

For ease of comparison with \DJetal{} we do not fit for $\sigma_\mathrm{instrument}$ but fix its value and do not include a contribution to the total error from the emulator in the likelihood. As previously discussed, we do not include the UVLF in our analysis.

\subsection{KL divergence}
\label{sec:kl-calc}

In order to calculate the KL divergence between the \textsc{ARES} and \textsc{globalemu} posteriors, we need to be able to evaluate the log-probability on both distributions for a common set of parameters given that
\begin{equation}
    \mathcal{D}_\mathrm{KL} = \langle \log P \rangle_{\theta \sim P(\theta|D, \mathcal{M})} - \langle \log P_\epsilon \rangle_{\theta \sim P(\theta|D, \mathcal{M})}.
\end{equation}
We use normalising flows~(NFs) implemented with \textsc{margarine} \citep{Bevins2022margarine1, Bevins2023margarine2} to evaluate the log-probabilities $\log P$ and $\log P_\epsilon$ for samples $\theta \sim P(\theta|D, \mathcal{M})$. NFs are a class of generative density estimation tools that use neural networks to parameterize invertible transformations between a known distribution such as a multivariate standard normal and a more complex target distribution like $P$ and $P_\epsilon$. Once trained, the flows can be used to generate samples on the distributions, i.e. $\theta \sim P(\theta|D, \mathcal{M})$, and via the change of variables formula calculate log-probabilities. Errors on the KL estimates are calculated using the method detailed in \cite{Bevins2023margarine2}.

For each normalising flow used in this work, we have a learning rate of $10^{-3}$ and five neural networks chained together, each with one hidden layer of 250 nodes. We train for a maximum of 1000 epochs and early stop if the validation loss has not decreased after 20 epochs, rolling back to the optimum model. For more details on the particular implementation of normalising flows used in this work, see \cite{Bevins2022margarine1, Bevins2023margarine2}.

When estimating KL divergence from samples, as is done here, we note that it can be oversensitive to the tails of the distributions if they have been poorly sampled. Equally, if the support of $P_\epsilon$ does not overlap with the support of $P$ the KL divergence becomes numerically unstable and equal to infinity. We note that this does not affect the limit defined in \cref{eq:limit-dkl} only the numerical approximations made in the following sections to illustrate the utility of the limit. As previously discussed, in general the KL divergence between $P$ and $P_\epsilon$ is unattainable because the true posterior is too computationally expensive.

\subsection{Results}

\begin{figure*}
    \centering
    \includegraphics[width=\linewidth]{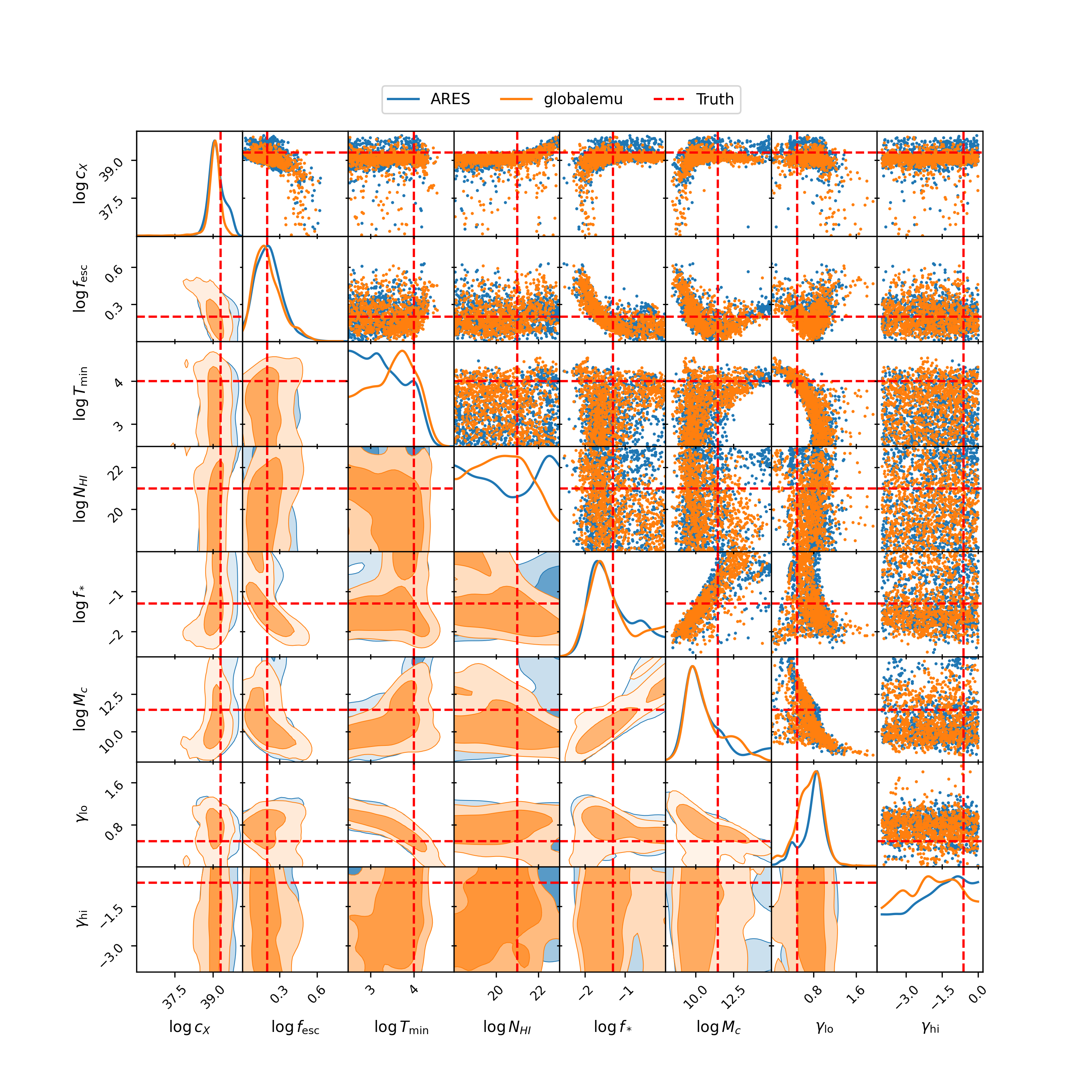}
    \caption{The posteriors recovered when fitting the fiducial \textsc{ARES} signal directly with \textsc{ARES} in blue and \textsc{globalemu} in orange for $\sigma=25$ mK. The lower half of the triangle plot shows a kernel density estimation~(KDE) of the 2D marginalised posteriors, the diagonal shows the 1D KDEs and the upper half shows the samples. We show the fiducial parameter values as red dashed lines for reference, but stress that we are more interested in the similarity between the posteriors in this work. While there are some small difference between the posteriors, they are visually similar. However, given the mean RMSE for the emulator and the limit outlined in \cref{eq:limit-dkl} this is not so surprising with the maximum predicted $\mathcal{D}_\mathrm{KL}$ between the emulated and true posteriors being $0.38$ bits. The actual $\mathcal{D}_\mathrm{KL}$ estimated with \textsc{margarine} is $0.05_{-0.52}^{+4.02}$.\label{fig:25mk-results}}
\end{figure*}

Using the trained emulator, the likelihood function with a fixed noise and \textsc{margarine} to calculate the KL divergence, as discussed in the previous sections, we are able to compare the posteriors recovered when using \textsc{ARES} and \textsc{globalemu} to model the signal. We use the nested sampling implementation \textsc{Polychord} \citep{Handely2015Polychord1, Handley2015Polychord2} to sample the likelihood function. In all the fits presented, we use the default \textsc{Polychord} settings\footnote{See \url{https://github.com/htjb/validating_posteriors}.} and we show in \cref{app:repeated-inference} that this is sufficient to recover consistent results. As in \DJetal{}, we test recovery when fitting a fiducial 21-cm signal with Gaussian distributed noise that has a standard deviation of 5, 25, 50 and 250 mK. We show the posteriors for 25 mK noise in \cref{fig:25mk-results} and for 5, 50 and 250 mK in \cref{app:additional-posteriors}.

From a visual inspection of the posterior distributions, we can see that even with $\sigma = 5$ mK the recovered posteriors are qualitatively similar when using \textsc{ARES} and \textsc{globalemu}. We plot the true values of the parameters used for the fiducial signal as a reference, but stress that the accuracy of the parameter estimation is less important here, and we are more concerned with the similarity between the recovered posteriors.

From \cref{fig:emulator-accuracy}, we might expect the recovery of the 1D posteriors on $f_\mathrm{esc}$, $C_X$ and $\log N_\mathrm{HI}$ to be worse than other parameters as the emulator error is worse at lower redshifts. However, we find that the posteriors on these parameters are very consistent, suggesting that the performance of the emulator even in this part of the parameter space is good. We report the `emulator bias' for each parameter and each level of noise in \cref{app:emulator-bias}. We find that the emulation bias is consistently below 1 for all the parameters and all the noise levels. At most, the emulation bias is 0.69 for $\log T_\mathrm{min}$ when $\sigma=5$ mK and we find that the average emulation bias decreases with increasing noise from 0.33 to 0.04 at 250 mK.

Using \cref{eq:limit-dkl} we can estimate an approximate upper limit on the $\mathcal{D}_\mathrm{KL}$ between the two posteriors for a given level of noise in the data and a given emulator error. In \cref{tab:dkl-values} we report these limits using the mean and 95th percentile RMSE values for the emulator for each level of noise considered in this paper. Since the 95th percentile RMSE is larger than the mean value and the limit is $\propto \mathrm{RMSE}^2$ it gives a more conservative limit on the accuracy of the emulated posterior. We also estimate the $\mathcal{D}_\mathrm{KL}$ between the posteriors using the method described in \cref{sec:kl-calc} and report these values in \cref{tab:dkl-values}. We see that for all the different noise levels the $\mathcal{D}_\mathrm{KL}$ are approximately consistent, within uncertainty, with 0 and consistent with the upper bounds from \cref{eq:limit-dkl}.

\begin{figure*}
    \centering
    \includegraphics[width=\linewidth]{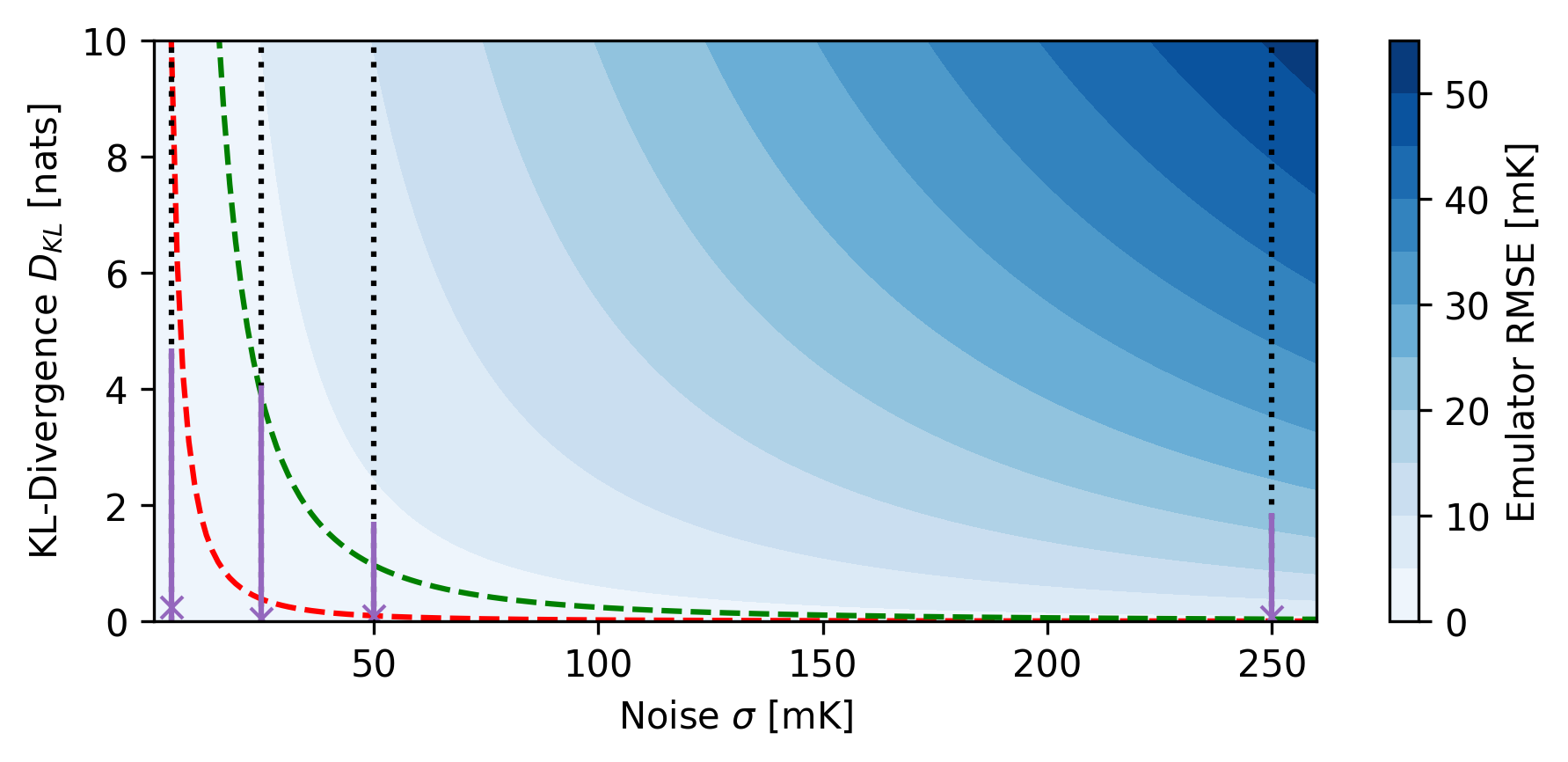}
    \caption{The graph shows how the upper limit value of $\mathcal{D}_\mathrm{KL}$, defined by \cref{eq:limit-dkl}, changes with the standard deviation of the Gaussian random noise in the data and the RMSE error on the emulator. The dashed red line and dashed green line show the contours corresponding to the mean and 95th percentile errors for the \textsc{globalemu} emulator used in this work. From the intersection between these lines and the dotted vertical lines at $\sigma=5, 25, 50$ and 250 mK one can put an approximate upper bound on the $\mathcal{D}_\mathrm{KL}$ between the posterior recovered when using \textsc{ARES} and the emulator. These upper bounds are reported in \cref{tab:dkl-values} with the bound from the 95th percentile being more conservative than from the mean RMSE across the test data. The purple scatter points show estimates of the KL divergence between the recovered posteriors for the three different noise levels. We see that even when $\sigma = 5$ mK the emulated posterior is very close to the true posterior recovered with \textsc{ARES} and that the upper limit defined in \cref{eq:limit-dkl} while not perfect provides a good gauge on the expected KL for a given emulator error. The KL values shown in purple are also reported in \cref{tab:dkl-values}. In this example, the units on $\sigma$ and RMSE are given in mK, but we stress that the discussion in this paper is applicable beyond 21-cm cosmology.\label{fig:kl-div}}
\end{figure*}

In \cref{fig:kl-div} we show how the upper limit on $\mathcal{D}_\mathrm{KL}$ changes with the magnitude of the noise in the data and the RMSE error on the emulator as a series of blue contours. We also show contours for the mean and 95th percentile RMSE values for the emulator used in this paper as red and green dashed lines. The vertical dotted lines mark the 5,  25, 50 and 250 mK noise values and the intersection between these lines and the dashed red and green lines give the predicted upper limits on the $\mathcal{D}_\mathrm{KL}$ reported in \cref{tab:dkl-values}. The purple scatter points give the $\mathcal{D}_\mathrm{KL}$ values, estimated with \textsc{margarine}, for each pair of posteriors. Although not perfect, it is clear that the limit in \cref{eq:limit-dkl} provides an approximate estimate for the maximum value of the $\mathcal{D}_\mathrm{KL}$ we might expect for a given noise and RMSE, and hence a good guide on the accuracy required of emulators for inference. For each pair of posteriors, the estimated $\mathcal{D}_\mathrm{KL}$ is approximately consistent with zero.

\begin{table}
    \centering
    \begin{tabular}{|c|c|c|c|}
    \hline
        Noise Level [mK] & \multicolumn{2}{c}{Estimated $\mathcal{D}_\mathrm{KL} \leq$} & Actual $\mathcal{D}_\mathrm{KL}$\\
        \hline
          & Mean RMSE & 95th Percentile & \\
          \hline
          5 & 9.60 & 96.62 & $0.25^{+4.45}_{-0.25}$\\
          25 & 0.38 & 3.86 & $0.05^{+4.02}_{-0.52}$\\
          50 & 0.10 & 0.97 & $0.09^{+1.62}_{-0.03}$\\
          250 & 0.004 & 0.039 & $0.08^{+1.78}_{-0.02}$ \\
          \hline
    \end{tabular}
    \caption{The table shows the predicted upper limits on $\mathcal{D}_\mathrm{KL}$ for three different noise levels $\sigma = 5, 25, 50$ and 250 mK using the mean and 95$^\mathrm{th}$ percentile RMSE values for the emulator used in this work. The upper limits come from \cref{eq:limit-dkl} and we also report the $\mathcal{D}_\mathrm{KL}$ values estimated with the code \textsc{margarine}. We find that the actual $\mathcal{D}_\mathrm{KL}$ are largely consistent with the bounds and zero within error.\label{tab:dkl-values}}
\end{table}

Since we used the same fiducial signal as \DJetal{} we should be able to visually compare the posteriors recovered with \textsc{ARES}. The comparison is complicated by the inclusion of constraints from the UV luminosity function in \DJetal{}, which will constrain the parameters governing star formation, but we can focus on the four parameters that are largely constrained by the sky-averaged 21-cm signal namely $f_\mathrm{esc}$, $C_X$, $T_\mathrm{min}$ and $\log N_\mathrm{HI}$. There is a similarity between the posteriors for the 5 mK noise in our work and the 25 mK noise in \DJetal{}. Similarly, consistent posteriors are seen on these four parameters when comparing our 50 mK case with the \DJetal{} 250 mK case. \DJetal{} shows the posteriors for the 50 mK case without including constraints from the UVLF allowing us to compare the posteriors for all the parameters in this case. Again, we see a similarity between our 25 mK case and the \DJetal{} 50 mK case. In the \DJetal{} 250 mK case, the 1D posteriors for $f_\mathrm{esc}$ and $T_\mathrm{min}$ appear to be quite well constrained, and they place an upper bound on the value of $C_X$. We do not see as strong constraints in our 250 mK posteriors (see \cref{app:additional-posteriors}).

\section{Discussion and Conclusions}
\label{sec:conclusions}

The upper bound on the KL divergence given in \cref{eq:limit-dkl} can be used to estimate how accurate a neural network emulator needs to be for accurate posterior recovery. While we have motivated this with an example in 21-cm cosmology, emulators are being widely used in a variety of fields and the discussion outlined here is more widely applicable. 

\DJetal{} concludes, based on the `emulator bias' in \cref{eq:emulator-bias}, that the recovered posteriors are inconsistent when using \textsc{ARES} and \textsc{globalemu} for data $\sigma \leq 25$ mK. However, we find that consistent constraints can be recovered using \textsc{globalemu} even for $\sigma = 5$ mK and frame the comparison in terms of the KL divergence. A similar level of agreement was found in \cite{Breitman202421cmEMU} when comparing the posteriors recovered from an analysis of the HERA 21-cm power spectrum upper limit with the emulator \textsc{21cmEMU} and the semi-numerical simulation code \textsc{21cmFAST}. Direct comparison with \DJetal{} is challenging because they included the UVLF in their analysis. The analysis and code from this paper is available at \url{https://github.com/htjb/validating_posteriors}.

The authors of \DJetal{} recently introduced a new emulator to the field, named \textsc{21cmLSTM}. This emulator, based on Long Short-Term Memory Neural Networks \citep{DorigoJones202421cmLSTM}, represents a significant improvement in accuracy over the current state-of-the-art emulators. They do not repeat the analysis done in \DJetal{} but state that an `emulation error $< 1$ mK is needed to sufficiently exploit optimistic or standard measurements of the 21-cm signal and obtain unbiased posteriors'. While similar in its heuristic nature to previous claims about required levels of accuracy \cite[e.g. that in ][]{Bevins2021globalemu}, this claim may be influenced by the results presented in \DJetal{}. This work provides a theoretical way to motivate the required level of emulator accuracy needed to recover an unbiased posterior estimate. The theoretical estimate supports the intuition laid out in \cite{Bevins2021globalemu} that the error in the emulator should be $\lesssim 10 \%$ of the expected noise in the data.

Although this may not have been the intention, \DJetal{} raised some questions about the constraints on the properties of the first galaxies and the early universe derived using \textsc{globalemu} from a number of works \citep[e.g.][]{Bevins2022SARAS2, Bevins2022SARAS3, Bevins2024Constraints, Pochinda2024constraints, GesseyJones2024constraints} but also works that have used other emulators such as \textsc{21cmGEM} \citep[e.g.][]{EDGESHB2019Constraints} and power spectrum emulators \citep[e.g.][]{HERA2022Constraints, HERA2023Constraints}. The theoretical arguments laid out in this paper, the experimental results and the availability of the code associated with this work should reaffirm confidence in these constraints and indeed in our ability to use emulators in 21-cm cosmology and beyond.

\section*{Acknowledgements}

We would like to thank the authors of \DJetal{} for providing the training and validation data used in their paper, and for providing comments on an early draft of this manuscript. We would also like to thank Jiten Dhandha for providing a curated list of 21-cm experiments at \url{https://github.com/JitenDhandha/21cmExperiments}.

HTJB acknowledges support from the Kavli Institute for cosmology Cambridge and the Kavli Foundation. WJH thanks the Royal Society for their support through their University Research Fellowships. TGJ acknowledges the support of the Science and Technology Facilities Council (UK) through grant ST/V506606/1 and the Royal Society.

This work used the DiRAC Data Intensive service (CSD3, project number ACSP289) at the University of Cambridge, managed by the University of Cambridge University Information Services on behalf of the STFC DiRAC HPC Facility (www.dirac.ac.uk). The DiRAC component of CSD3 at Cambridge was funded by BEIS, UKRI and STFC capital funding and STFC operations grants. DiRAC is part of the UKRI Digital Research Infrastructure.

%%%%%%%%%%%%%%%%%%%%%%%%%%%%%%%%%%%%%%%%%%%%%%%%%%
\section*{Data Availability}

The code used in this paper is publicly available at \url{https://github.com/htjb/validating_posteriors} and the training and test data are available on Zenodo at \url{https://doi.org/10.5281/zenodo.15040279}.

%%%%%%%%%%%%%%%%%%%% REFERENCES %%%%%%%%%%%%%%%%%%

% The best way to enter references is to use BibTeX:

\bibliographystyle{mnras}
\bibliography{example} % if your bibtex file is called example.bib

% Alternatively you could enter them by hand, like this:
% This method is tedious and prone to error if you have lots of references
%\begin{thebibliography}{99}
%\bibitem[\protect\citeauthoryear{Author}{2012}]{Author2012}
%Author A.~N., 2013, Journal of Improbable Astronomy, 1, 1
%\bibitem[\protect\citeauthoryear{Others}{2013}]{Others2013}
%Others S., 2012, Journal of Interesting Stuff, 17, 198
%\end{thebibliography}

%%%%%%%%%%%%%%%%%%%%%%%%%%%%%%%%%%%%%%%%%%%%%%%%%%

%%%%%%%%%%%%%%%%% APPENDICES %%%%%%%%%%%%%%%%%%%%%

\appendix

\section{Additional Results}
\label{app:additional-posteriors}

\begin{figure*}
    \centering
    \includegraphics[width=\linewidth]{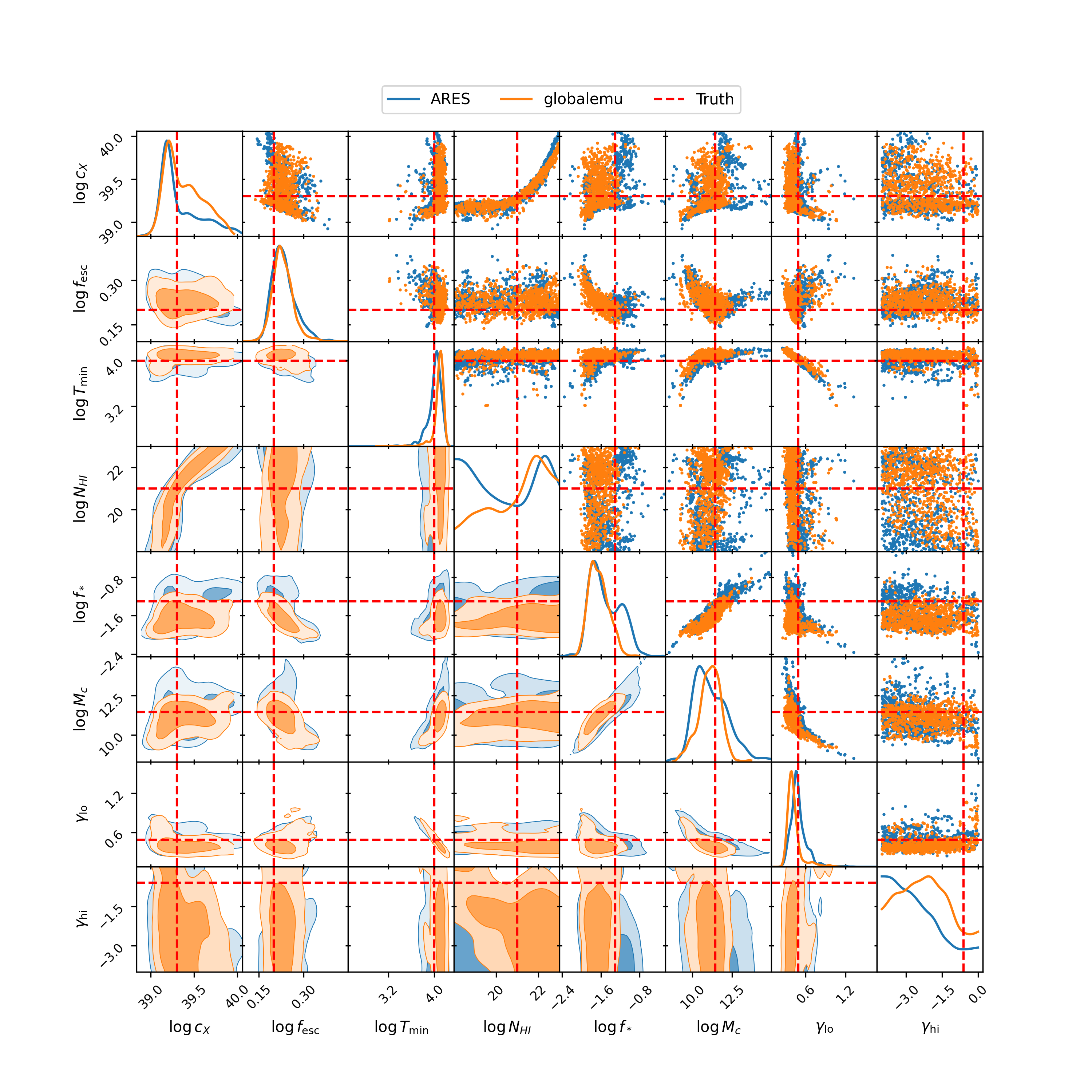}
    \caption{The figure shows the recovered posteriors when modelling the data directly with \textsc{ARES} in blue and with \textsc{globalemu} in orange for $\sigma=5$ mK. As with \cref{fig:25mk-results} there is a similarity between the two posteriors and although the upper limit on the $\mathcal{D}_\mathrm{KL} = 9.60$ bits the calculated $\mathcal{D}_\mathrm{KL}$ is $0.25_{-0.25}^{+4.45}$.\label{fig:5mk-results}}
\end{figure*}

\begin{figure*}
    \centering    \includegraphics[width=\linewidth]{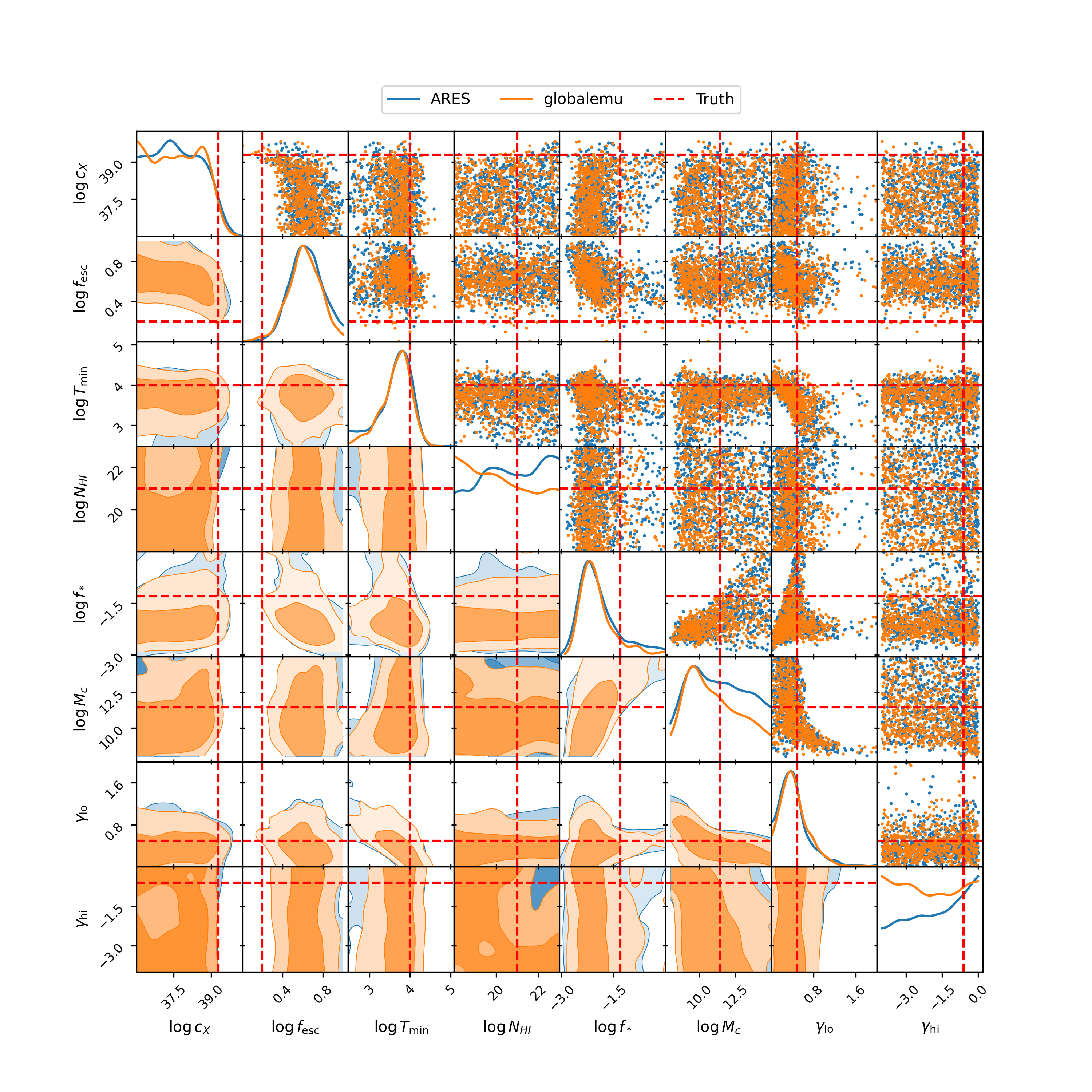}
    \caption{The true posterior recovered with \textsc{ARES} in blue and in orange posterior recovered with the \textsc{globalemu} emulator for $\sigma=50$ mK. As expected from \cref{eq:limit-dkl} and \cref{fig:kl-div} the posterior distributions look even more alike than when the noise is 5 and 25 mK. The estimated $\mathcal{D}_\mathrm{KL} \leq 0.10$ based on the mean emulator RMSE or $\mathcal{D}_\mathrm{KL} \leq 0.97$ based on the 95th percentile emulator RMSE. The calculated $\mathcal{D}_\mathrm{KL} = 0.09_{-0.03}^{+1.62}$.\label{fig:50mk-results}}
\end{figure*}

\begin{figure*}
    \centering    \includegraphics[width=\linewidth]{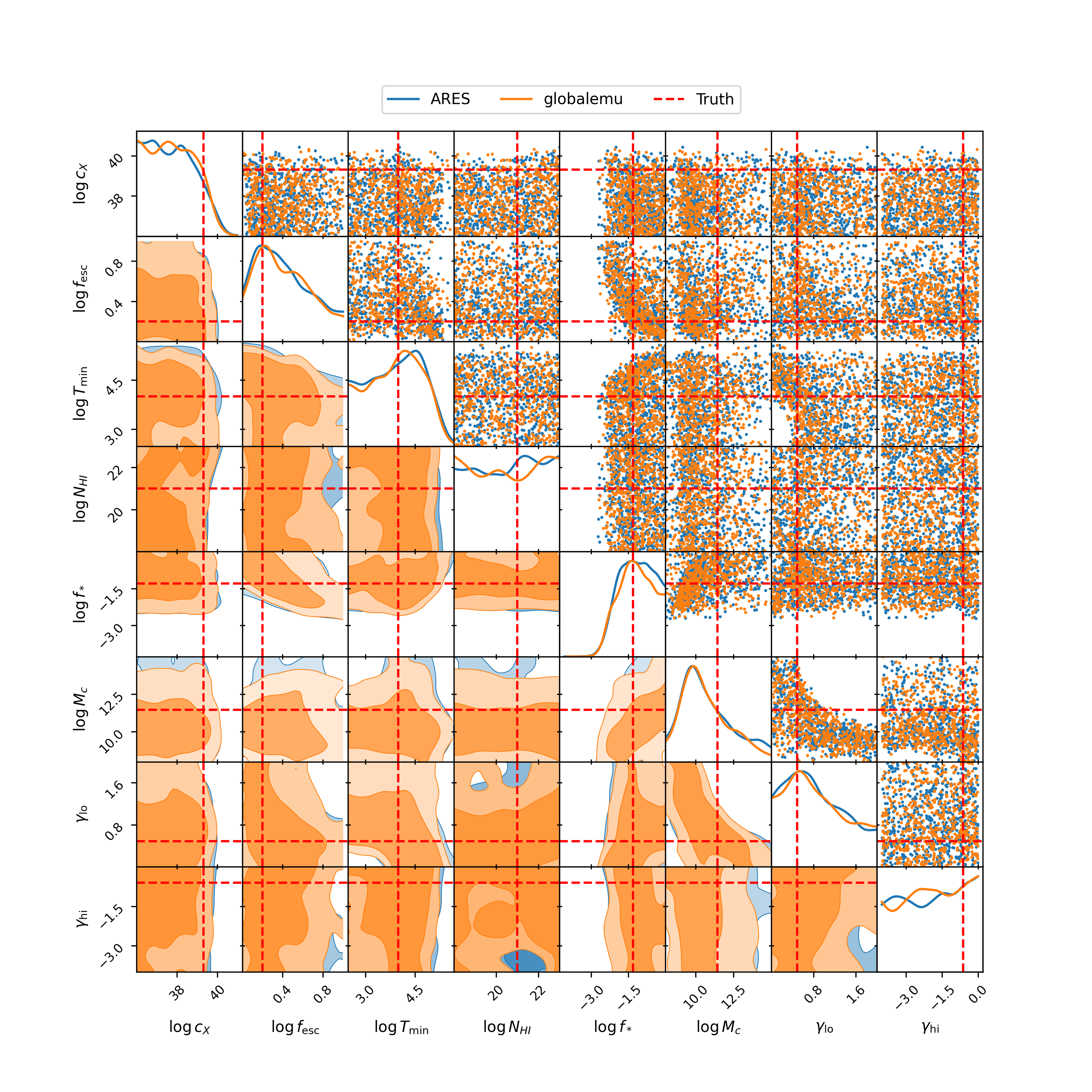}
    \caption{The true posterior recovered with \textsc{ARES} in blue and in orange the posterior recovered with \textsc{globalemu} for $\sigma=250$ mK. The estimated upper limit on the $\mathcal{D}_\mathrm{KL}$ for this level of noise is 0.004 nats compared with the calculated value of $\mathcal{D}_\mathrm{KL} = 0.08_{-0.02}^{+1.78}$.\label{fig:250mk-results}}
\end{figure*}

In \cref{fig:5mk-results}, \cref{fig:50mk-results} and \cref{fig:250mk-results} we show the true and emulated posteriors recovered when $\sigma=5, 50$ and 250 mK. For $\sigma=5 $ mK we are able to accurately recover the true posterior when using \textsc{globalemu}.

\section{Emulator Bias}
\label{app:emulator-bias}

\begin{figure*}
    \centering
    \includegraphics[width=\linewidth]{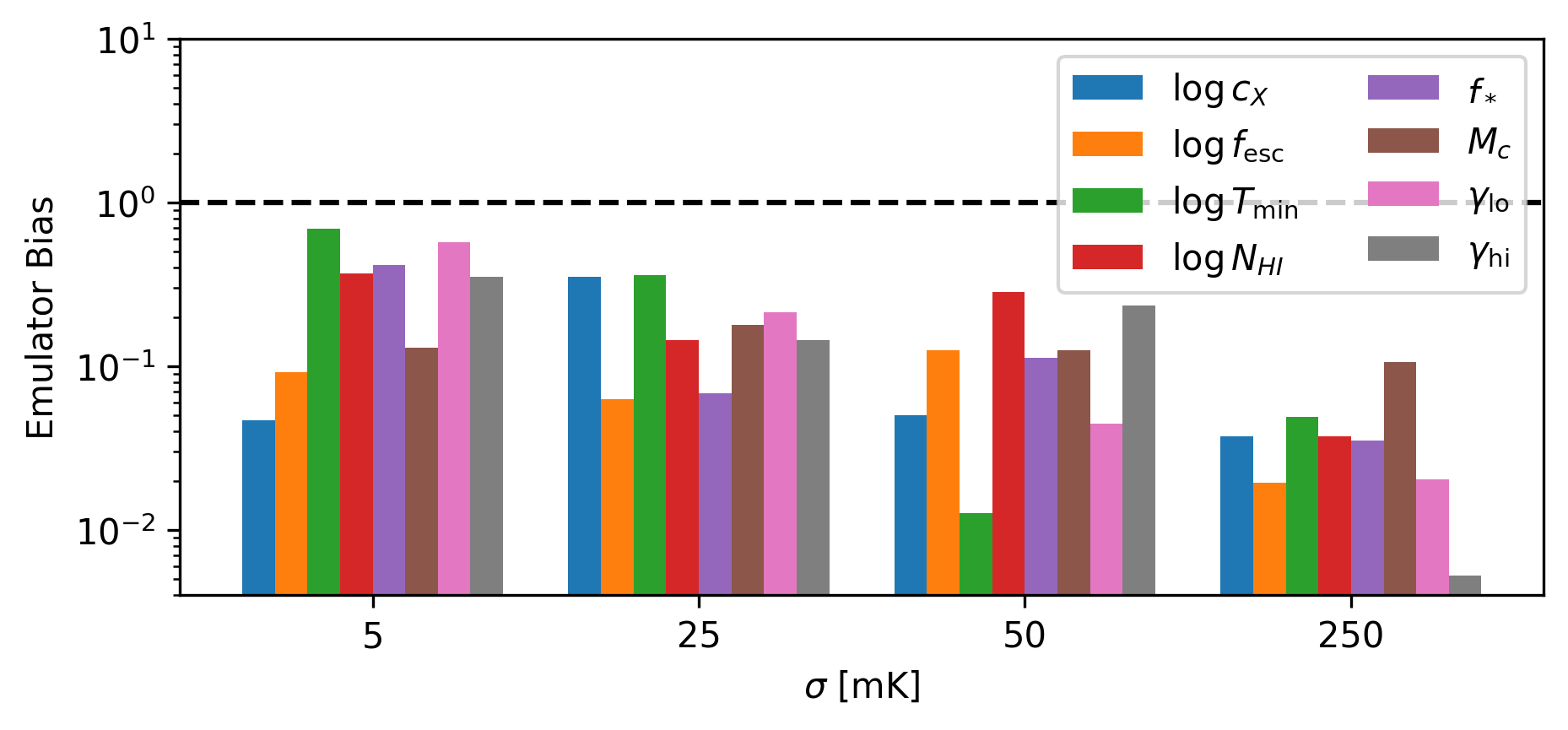}
    \caption{The plot shows the Emulator Bias, as defined in \cref{eq:emulator-bias}, for each parameter at each noise level. \DJetal{} suggests that an emulator bias $> 1$ indicates a poor recovery of the posterior. \DJetal{} finds that for $\sigma \leq 25$ mK the emulator bias begins to exceed this limit for $\log T_\mathrm{min}$ and $\gamma_\mathrm{lo}$ in particular. While we find that the emulator bias increases with decreasing noise, on average, we find that it never exceeds a value of 1 for any of the parameters.\label{fig:bias-comparison}}
\end{figure*}

In \cref{fig:bias-comparison} we show the `emulator bias' as defined by \cref{eq:emulator-bias} for each parameter and each noise level. The average bias increases with decreasing noise level, as we would expect, from 0.04 for 250 mK to 0.33 at 5 mK. The maximum bias at $\sigma=5$ mK is 0.69 which is less than the threshold for accurate posterior recovery defined in \DJetal{} of 1.

\section{Reproducibility of results}
\label{app:repeated-inference}

Nested sampling is designed to evaluate the Bayesian evidence $\mathcal{Z}$ and in the process return samples from the posterior distribution. The error in the nested sampling algorithm is determined by the number of samples evaluated on the likelihood. If this number is too small, then the posterior will not be fully resolved and the evidence will likely be underestimated.

The number of samples attained during a nested sampling run is largely governed by the number of live points that are evolved up the likelihood contours. If there are too few live points, then important features on the posterior can be missed and inconsistent results are recovered on repeated sampling. This is particularly true for multimodal posteriors. 

\begin{figure*}
    \centering
    \includegraphics[width=\linewidth]{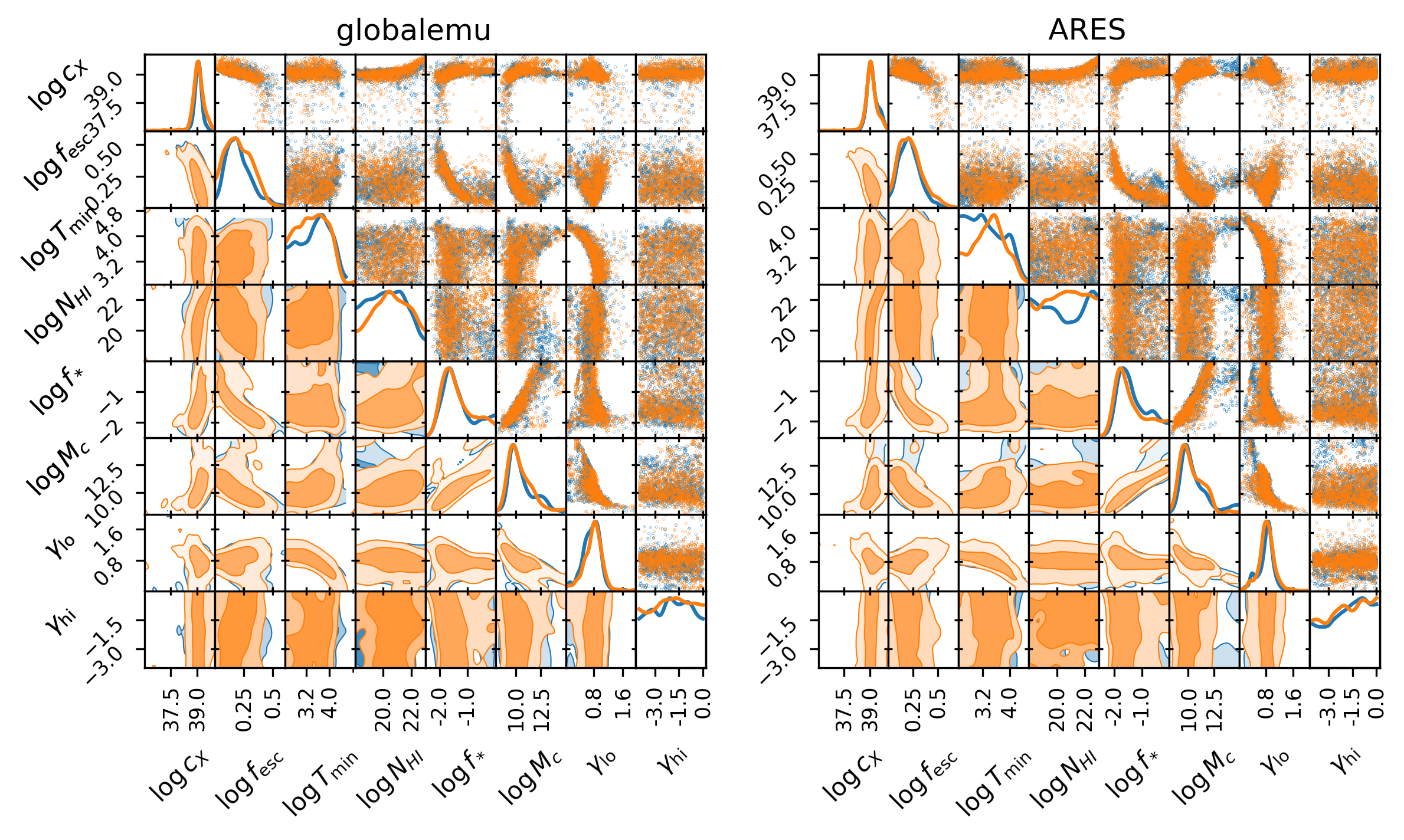}
    \caption{\textbf{Left Panel:} We show two sets of chains from running \textsc{polychord} with $n_\mathrm{live}=200$ on the \textsc{globalemu} fit to the fiducial \textsc{ARES} signal with $\sigma=25$ mK. The chains are visually similar and the Bayesian evidences for both fits are consistent within the sampling error as shown in \cref{tab:repeat-evidence}. \textbf{Right Panel:} The same as the left panel, but fitting the fiducial \textsc{ARES} signal plus noise directly with \textsc{ARES}. Again the posteriors appear consistent, and the evidences are consistent with error.}
    \label{fig:repeated-posteriors}
\end{figure*}

In our analysis, we use the default number of live points recommended in the \textsc{polychord} implementation of nested sampling, corresponding to $n_\mathrm{live} = 200$ or 25 per dimension. We rerun our analysis on the fiducial mock data with 25 mK to check that $n_\mathrm{live}$ is high enough when using both the emulator and \textsc{ARES}.
The results are shown in \cref{fig:repeated-posteriors}. We find that 200 live points is enough to recover consistent chains on repeated runs and report the corresponding Bayesian evidences in \cref{tab:repeat-evidence}.

\begin{table}
    \centering
    \begin{tabular}{|c|c|c|}
        & \multicolumn{2}{c}{$\log Z$} \\
        \hline
        & \textsc{globalemu} & \textsc{ARES} \\
        \hline
         Run 1 & $-2297.18 \pm 0.25$ & $-2297.01 \pm 0.25$ \\
         Run 2 & $-2297.26 \pm 0.25$ & $-2296.94 \pm 0.25$\\
         \hline
    \end{tabular}
    \caption{The Bayesian evidence for the four fits shown in \cref{fig:repeated-posteriors}. We repeat the inference on the fiducial \textsc{ARES} signal with 25 mK noise with both \textsc{globalemu} and \textsc{ARES} to check that the sampler, \textsc{polychord}, is set up appropriately such that we recover consistent chains. \cref{fig:repeated-posteriors} shows that the posteriors are visually the same and the table here shows that the evidences are consistent.}
    \label{tab:repeat-evidence}
\end{table}

%%%%%%%%%%%%%%%%%%%%%%%%%%%%%%%%%%%%%%%%%%%%%%%%%%

% Don't change these lines
\bsp	% typesetting comment
\label{lastpage}
\end{document}